\documentclass[review]{elsarticle}
\usepackage{lineno,hyperref}
\usepackage[english]{babel}
\usepackage[numbers]{natbib}
\usepackage{xcolor}
\usepackage{latexsym,amsmath,amssymb,amsbsy,graphicx,geometry}
\usepackage{epstopdf}
\usepackage{epsfig}
\modulolinenumbers[5]

\begin{document}
	
	\begin{frontmatter}
	
\title{Excellent glass former Ni$_{62}$Nb$_{38}$ crystallizing under combined shear and ultra-high pressure}

\author[kfu,ufrc]{Bulat N. Galimzyanov\corref{cor1}}
\cortext[cor1]{Corresponding author}
\ead{bulatgnmail@gmail.com}

\author[kfu]{Maria A. Doronina}
\ead{maria.doronina.0211@gmail.com}

\author[kfu,ufrc]{Anatolii V. Mokshin}
\ead{anatolii.mokshin@mail.ru}

\address[kfu]{Kazan Federal University, 420008 Kazan, Russia} 
\address[ufrc]{Udmurt Federal Research Center of the Ural Branch of the Russian Academy of Sciences, 426067 Izhevsk, Russia}

\begin{abstract}
Study of condensed matter in certain extreme conditions allows one to better understand the mechanisms of microscopic structural transformations and to develop materials with completely new mechanical properties. In the present work, we study the structure and crystallization kinetics of bulk metallic glass (BMG) Ni$_{62}$Nb$_{38}$. We show that this BMG is rapidly crystallized under the combined effect of shear deformation and ultrahigh pressure. A threshold pressure required to initiate the formation of a stable crystalline phase is revealed. Shear deformation and pressure lead to the phase separation: two high-density inhomogeneous crystal structures are formed. We find that the crystallization of this BMG occurs in two stages due to the significant difference in the growth rates of Ni and Nb crystalline phases. The results of the present study make a significant contribution to understanding the crystallization and amorphization features of Ni-based BMG's.
\end{abstract}


\end{frontmatter}

\section{Introduction}
The study of crystallization mechanisms of bulk metallic glasses (BMG's) with outstanding glass-forming ability becomes more available to the improvement of high-pressure devices~\cite{Dubrovinskaia_2016,Zhang_Ma_2017,Burrage_Perreault_2019,Glezer_2019}. Modern experimental methods make it possible to obtain static pressures more than $1000$~GPa, for example, in two-stage high-pressure devices with diamond anvils~\cite{Dubrovinsky_2015,Dewaele_Loubeyre_2018,Brazhkin_2020}. Such terapascal pressures comparable with pressures in Jupiter's interior are usually achieved for micron-size samples. Moreover, external pressure is an important factor for controlling the microscopic structure of BMG's that is confirmed experimentally~\cite{Xing_Jiang_2002,Brazhkin_Tsiok_2017} and based on the results of molecular dynamics simulations~\cite{Sosso_2016,Mo_Liu_2017}: ultrahigh pressure can both promote the crystallization of BMG's~\cite{Mo_Liu_2017,Watanauki_Shobu_2006,Galimzyanov_Mokshin_2018,Galimzyanov_AM_2019} and suppress this process~\cite{Soignard_Amin_2008,Wang_Yao_2001,Soignard_Brazhkin2020}. 

The impact of pressure on BMG structure strongly depends on the alloy composition and on the size of a sample. Most BMG's consist of two or more chemical elements. The difference in the atomic radii of these elements leads to the formation of a relatively stable amorphous structure~\cite{Halim_Ma_2021,Li_Adv_2020}. Among binary BMG's, the best glass-forming ability is observed in Ni-Nb system with composition Ni$_{62}$Nb$_{38}$~\cite{Xia_Li_2006,Suryanarayana_2018,Massalski_Okamoto_1990,Lesz_Dercz_2016}. Xia and co-workers have shown that samples of this BMG practically do not crystallize under normal conditions if their linear size is less than $2$~mm~\cite{Xia_Li_2006}. This feature distinguishes Ni$_{62}$Nb$_{38}$ from other binary Ni-based BMG's. 

The mechanisms of microscopic structural transformation in BMG Ni$_{62}$Nb$_{38}$ are poorly studied experimentally due to difficulties associated with monitoring the nucleation and growth of nanosized crystallites inside a bulk sample~\cite{Galimzyanov_AM_2019,Liu_Chen_2008,Carter_Williams_2009,Ediger_Harrowell_2012}. The initial stage of crystallization is difficult to track using X-ray and neutron spectroscopy data since these methods, as a rule, provide averaged information about the structure~\cite{Gleiter_2000,Myerson_2002}. On the other hand, until recently, detailed molecular dynamics simulations of the crystallization process in Ni$_{62}$Nb$_{38}$ were not performed due to the absence of interparticle interaction potentials capable of correctly reproducing the amorphous structure and physical properties of this system for a wide range of temperatures and pressures. A better understanding of structural ordering mechanisms on the atomistic scale becomes more available after the development of a semiempirical potential by Zhang and co-workers~\cite{Kelton_2016}. This potential accurately reproduces the bond angle distribution in Ni$_{62}$Nb$_{38}$ and allows one to obtain BMG with a relatively stable amorphous structure. Note that the bond-angle is the angle between the vectors joining a central particle with two other atoms. The bond-angle distributions are used to represent the local angular correlations in disordered media~\cite{Canales_Padro_1992}.

In the present work, we carry out the non-equilibrium molecular dynamics simulation study of the microscopic structure of metallic glass Ni$_{62}$Nb$_{38}$ crystallizing under the impact of external forces. We will show that the ``explosive'' crystallization of this BMG is observed only when the system undergoes a fixed ultrahigh pressure and shear deformation with a fixed shear rate.

\section{Computational procedure}

The considered system Ni$_{62}$Nb$_{38}$ consists of $13\,203$ Ni atoms and $8\,093$ Nb atoms (the total number of atoms is $N=21\,296$). The atoms are located inside a triclinic simulation box with the linear size $L_{x}=L_{y}=L_{z}\simeq67$~\AA. The interatomic interaction energies and forces are given by the semiempirical Finnis-Sinclair potential~\cite{Kelton_2016,Finnis_Sinclair_1984}: 
\begin{equation}\label{eq_FS_1}
U(r_{ij})=\sum_{i=1}^{N-1}\sum_{j\neq i}^{N}\phi_{\alpha_{i}\alpha_{j}}(r_{ij})+\sum_{i=1}^{N}\Phi_{\alpha_{i}}(\rho_{i}),
\end{equation}
where
\begin{equation}\label{eq_FS_2}
\rho_{i}=\sum_{j\neq i}^{N}\psi_{\alpha_{i}\alpha_{j}}(r_{ij}).
\end{equation}
Here, $\rho_{i}$ is the local electronic charge density at site $i$, which is constructed by a rigid superposition of atomic charge densities $\psi_{\alpha_{i}\alpha_{j}}(r_{ij})$, where $\alpha\in\{$Ni,\,Nb$\}$ is the type of atom; $r_{ij}$ is the distance between the atoms $i$ and $j$; $\phi_{\alpha_{i}\alpha_{j}}(r_{ij})$ is the pairwise potential; $\Phi_{\alpha_{i}}(r_{ij})$ is the embedding energy function. Note that this type of potential reproduces accurately the structure of Ni$_{62}$Nb$_{38}$ in the liquid and amorphous states~\cite{Kelton_2016}.

\begin{figure*}[ht]
	\centering
	\includegraphics[width=15.0cm]{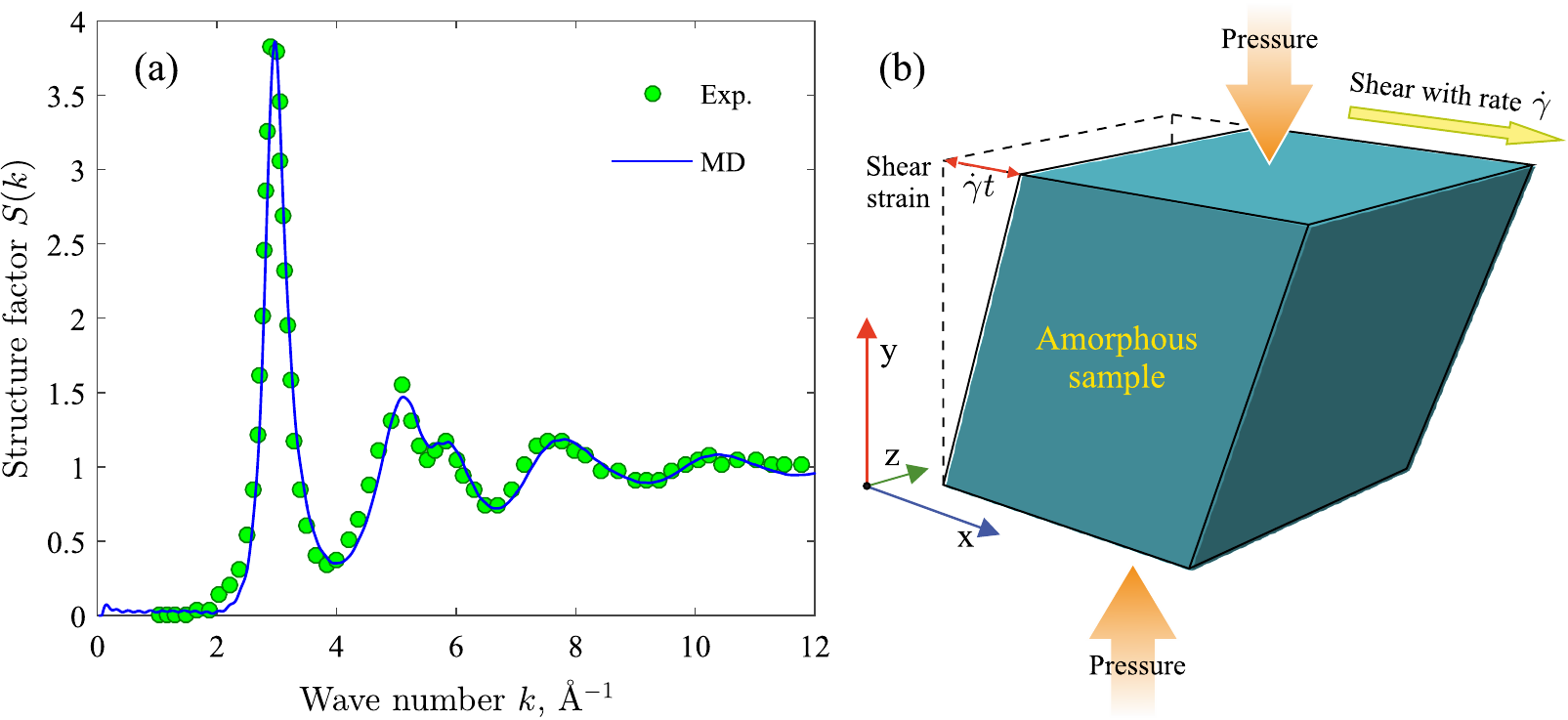}
	\caption{(a) Structure factor of the amorphous solid found from the molecular dynamics simulations (MD) at the temperature $300$~K and at the pressure $10^{-4}$~GPa is compared with the experimental data (Exp) from Ref.~\cite{Xu_Jiang_2017}. (b) Schematic representation of shear deformation of an amorphous sample.}\label{fig_1}
\end{figure*}

To prepare amorphous samples, liquid samples with the temperature $T=2500$~K are rapidly cooled to the temperature $T=300$~K at the pressure $p=10^{-4}$~GPa with the cooling rate $10^{13}$~K/s. To reveal the effect of the cooling procedure on the crystallization kinetics of the system, additional studies were carried out at the cooling rate $10^{10}$~K/s. Note that the liquidus and glass transition temperatures of Ni$_{62}$Nb$_{38}$ at zero pressure are $T_{l}\simeq1450$~K (the experimental value is $\approx1400$~K) and $T_{g}\simeq1010$~K, respectively, and these temperatures follow from the results of molecular dynamics simulations with the Finnis-Sinclair interaction potential~\cite{Kelton_2016}. Figure~\ref{fig_1}(a) shows that the structure factor $S(k)$ of amorphous Ni$_{62}$Nb$_{38}$ calculated at the temperature $T=300$~K is in agreement with the experimentally measured diffraction pattern~\cite{Xu_Jiang_2017}.
Prepared amorphous samples undergo shear deformation in the $xy$ plane and in the $x$-direction at the fixed shear rate $\dot\gamma=0.01$~ps$^{-1}$ as shown in Figure~\ref{fig_1}(b). This strain rate is typical in molecular dynamics simulations and this rate is difficult to obtain in experiments~\cite{Galimzyanov_Mokshin_2018,Mokshin_Barrat_2013,Chang_Zhou_2017,Sha_2018}. For example, shear rates from $0.0001$~ps$^{-1}$ to $0.1$~ps$^{-1}$ were used by Chang and co-workers to study the effect of shear strain on the behavior of single-crystal titanium nanowire~\cite {Chang_Zhou_2017}. They have shown that the physicomechanical properties of titanium nanowire are weakly dependent on the strain at shear rates in the interval $\dot\gamma\in[0.0001;\,0.01]$~ps$^{-1}$. Sha and co-workers are achieved to the same conclusions in the study of single-layer borophene mechanical properties at shear rates from $10^{-5}$~ps$^{-1}$ to $0.01$~ps$^{-1}$~\cite{Sha_2018}. Thus, we reasonably expect that structural transformations occurring in metallic glass Ni$_{62}$Nb$_{38}$ at the shear rate $\dot\gamma=0.01$~ps$^{-1}$ will also be observed in experiments at lower shear rates. Here, it should be noted that this relationship between simulation and experiment may not be fulfilled in the case of systems with complex chemical bonds. We perform shear deformation at the fixed pressures $p=200$, $400$, $600$, $800$, $1000$~GPa and these pressures are experimentally achievable with high-pressure devices with diamond anvils~\cite{Dubrovinskaia_2016,Dubrovinsky_2015,Dewaele_Loubeyre_2018,Brazhkin_2020}. These pressures are applied only along the $y$ direction -- the so-called gradient direction [see Figure~\ref{fig_1}(b)]. The constant pressure $10^{-4}$~GPa acts only in the $z$-direction.  

Bond orientational analysis is performed for the identification of different crystalline phases and clusters inside amorphous samples. The local $q_l$ and the global $Q_l$ orientational order parameters (here $l=4,\,6,\,8$) are computed according to the definitions given in Refs.~\cite{Steinhardt_1983,Wolde_Frenkel_1995}. To identify the crystalline phase atoms, we compute the local orientational order parameters $q_{l}$ separately for Ni and Nb atoms [see Eq.~(\ref{eq_A1}) in Appendix]. The existence of atoms with values of $q_l$ close to those of an ideal ordered fcc, hcp, bcc, simple cubic, and icosahedron structures is evidence of the presence of ordered clusters [see Table I in Ref.~\cite{Mickel_Kapfer_2013}]. To detect a single crystalline order across the entire sample, we compute the global orientational order parameters $Q_{l}(\alpha)$ (here $\alpha=\{$Ni$,\,$Nb$\}$) [see Eq.~(\ref{eq_A5}) in Appendix]. Note that these order parameters with indices $l=4,\,6,\,8$ are sufficient to correctly detect ordered structures.

\section{Results and Discussions}

\begin{figure*}[ht]
	\centering
	\includegraphics[width=11.0cm]{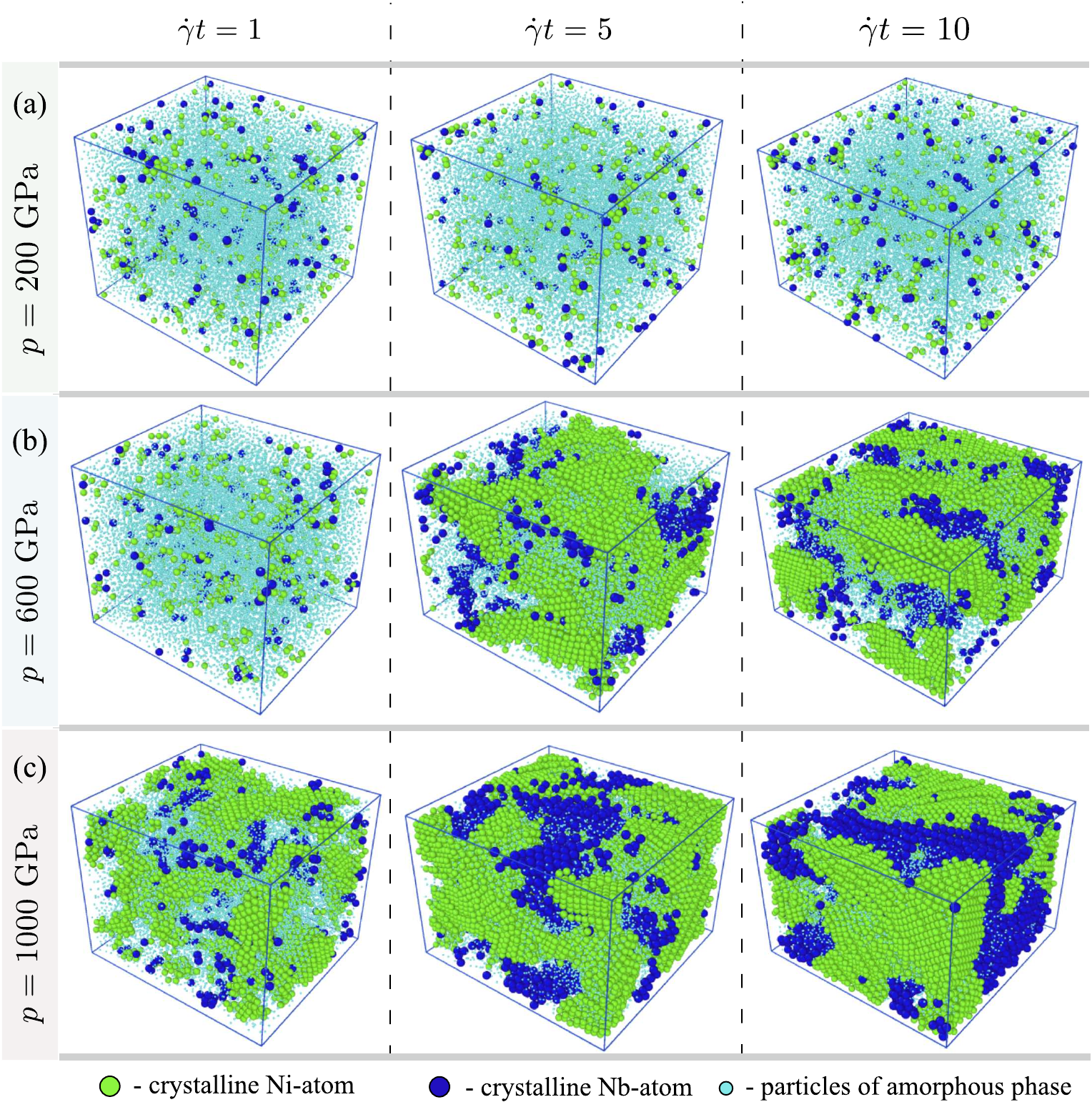}
	\caption{Crystallizing amorphous Ni$_{62}$Nb$_{38}$ at the different strains ($\dot\gamma t=1$, $5$, and $10$) and at the various pressures: (a) $p=200$~GPa; (b) $p=600$~GPa; (c) $p=1000$~GPa. Crystalline Ni and Nb atoms are marked by green and blue colors, respectively. The atoms of the parent disordered phase are colored by light blue.}\label{fig_2}
\end{figure*}

The impact of high pressure and shear initiates the phase transition in the amorphous alloy Ni$_{62}$Nb$_{38}$. We have found that intense formation and growth of stable crystalline nuclei occur only at pressures above $p=200$~GPa. Regardless the magnitude of the strain $\dot{\gamma}t$, the formation of stable crystallites is not observed at lower pressures. Figure~\ref{fig_2}(a) shows that unstable nanocrystallites randomly distributed throughout the amorphous matrix are formed in the system at $p\leq200$~GPa. The concentration of these nanocrystallites does not change with increasing strain up to $\dot{\gamma}t=10$. Such poor crystallizability may be due to the fact that the degree of supercooling of the system at the temperature $300$~K is $\Delta T/T_{l}\simeq0.79$ (where $\Delta T=T_{l}-T$). At this supercooling, the mobility of atoms is suppressed by high viscosity~\cite{Galimzyanov_Mokshin_2018,Mokshin_Galimzyanov_2017}. The energy entered into the glassy system due to applied pressure and shear is insufficient to relax this system into the equilibrium crystalline phase.  

``Explosive'' crystallization of Ni$_{62}$Nb$_{38}$ occurs at pressures from $600$ to $1000$~GPa. At such pressures, the crystallization degree of the system increases in a direct proportion to the strain $\dot{\gamma}t$ [see Figures~\ref{fig_2}(b) and~\ref{fig_2}(c)]. The higher the applied pressure, the less shear strain is required to initiate the crystallization process. This is clearly seen from Figure~\ref{fig_2}(c), where the sample with crystalline domains is formed at the pressure $p=1000$~GPa and at strains $\dot{\gamma}t\leq1$, whereas crystallization is not observed at similar strains and at the pressure $p=600$~GPa [see Figure~\ref{fig_2}(b)]. This finding is of great practical importance, since crystalline samples obtained at such relatively small shear strains do not experience pronounced mechanical injuries~\cite{Zhang_Duan_2019}.

Crystalline nuclei with a non-spherical shape and consisting of several tens of atoms are formed at the initial stage of BMG crystallization. The non-spherical shape of these nuclei is due to the impact of shear strain as it was shown earlier, for example, in Refs~\cite{Mokshin_Barrat_2013,Mura_Zaccone_2016,Galimzyanov_JR_2018}. The formation of these crystal nuclei obeys the Poisson distribution, and this observation is in good agreement with the main postulates of the classical theory of homogeneous stationary nucleation presented in Refs. ~\cite{Kashchiev_2000,Kelton_Greer_2010,Kalikmanov_2012,Schmelzer_Abyzov_2018}. However, further growth of crystallites does not follow the classical scenario~\cite{Karthika_Kalaichelvi_2016}. We have found that shear strain leads to crystal separation and formation of two fractions with different composition and structure. One of the fractions consists only of Ni atoms, whereas the second one contains of Nb atoms. Figure~\ref{fig_2} shows that this separation is clearly manifested at high pressures (over $400$~GPa) and at deformations $\dot{\gamma}t>1$. The growth of each fraction occurs both due to the attachment of individual Ni or Nb atoms and due to the coalescence of small crystallites. The coalescence of nuclei leads to the formation of the inhomogeneous crystal structure typical for phase separation~\cite{Skripov_1979,Igolnikov_Skripov_2021}. Note that a similar diffusive redistribution of components with a change in the crystal structure is observed in many multicomponent alloys undergoing phase transformations~\cite{Yang_Tang_2020,Rudraraju_2016,Bachmaier_2015}. For example, Bachmaier et al. have found the spinodal-type decomposition and formation of a nanoscale composite structure in the binary alloy Cu-Co under high-pressure torsion deformation~\cite{Bachmaier_2015}. Yang and Ming have observed surface-localized phase separation behavior of LiFePO4 (cathode material for Li-ion batteries) induced by elastic deformation and lattice expansion due to Li insertion~\cite{Yang_Tang_2020}.  
\begin{figure*}[ht]
	\centering
	\includegraphics[width=11.0cm]{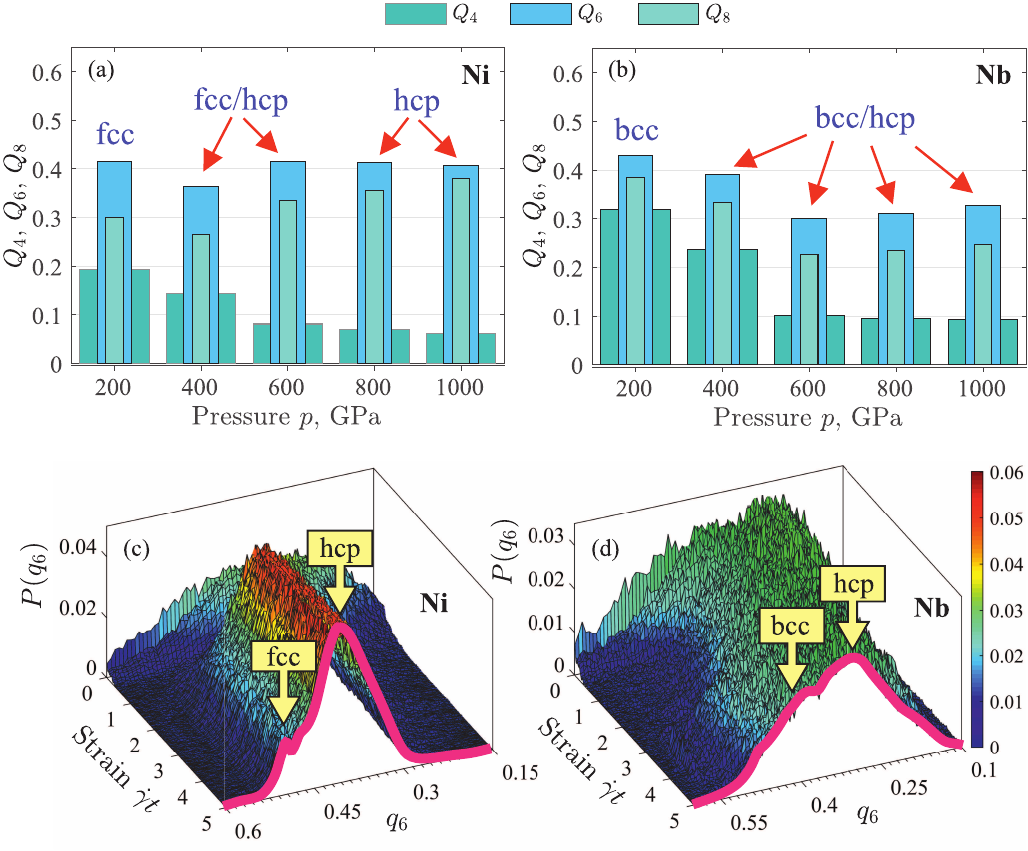}
	\caption{Global orientational order parameters $Q_{4}$, $Q_{6}$, $Q_{8}$ at different pressures and at the strain $\dot{\gamma}t=5$: (a) for the crystal structure of Ni; (b) for the crystal structure of Nb. Note that the values of these parameters were calculated without applying the correlation condition (\ref{eq_A3}). 3D-distribution of the atoms $P(q_{6})$ over the value of the order parameter $q_{6}$ is presented as a function of the strain $\dot{\gamma}t$ at $p=600$~GPa: (c) for Ni atoms; (d) for Nb atoms. Arrows indicate the type of crystal lattice: fcc (face-centered cubic), hcp (hexagonal closest packed), and bcc (body-centered cubic).}\label{fig_3}
\end{figure*}

From the found values of the global orientational order parameters $Q_{4}$, $Q_{6}$ and $Q_{8}$ [see. Figures~\ref{fig_3}(a) and~\ref{fig_3}(b)] it follows that the nanocrystallites of Ni have the fcc lattice at the pressure $p=200$~GPa and at the strain $\dot{\gamma}t=5$, while Nb forms the bcc phase under these conditions. Note that at low pressures the fcc and bcc phases are typical for Ni and Nb, respectively~\cite{Sharma_Hickman_2018,Bollinger_White_2011}. The system forms a high-density crystalline phase close to hcp at pressures above $400$~GPa. In the case of Ni, the mixing of the hcp and fcc structures is observed, whereas Nb forms the mixed hcp-bcc phase. As an example, Figures~\ref{fig_3}(c) and~\ref{fig_3}(d) show the distribution $P(q_{6})$ of the order parameter $q_{6}$ calculated for Ni and Nb crystallites at different strains and at the pressure $p=600$~GPa. The shape of the distribution $P(q_{6})$ directly indicates the predominance of the number of atoms forming the hcp structure over the atoms entering the fcc and bcc structures. As can be seen from Figures~\ref{fig_3}(c) and~\ref{fig_3}(d), the shape of the 3D-distributions varies weakly with the degree of strain at $\dot{\gamma}t>1$. This finding indicates that the pressure (or density of the system) plays a crucial role in the formation of a type of crystal lattice symmetry during the crystallization of amorphous Ni$_{62}$Nb$_{38}$. It should be noted that the hcp, fcc, and bcc crystal lattices are distorted as a result of applied shear strain. Therefore, the calculated values of the parameters $Q_{l}$ ($l=4,\,6,\,8$) and the distribution $P(q_{6})$ [see Figure~\ref{fig_3}] may differ from the known literature data obtained for systems with an ideal crystal structure~\cite{Mickel_Kapfer_2013}.

\begin{figure*}[ht]
	\centering
	\includegraphics[width=11.0cm]{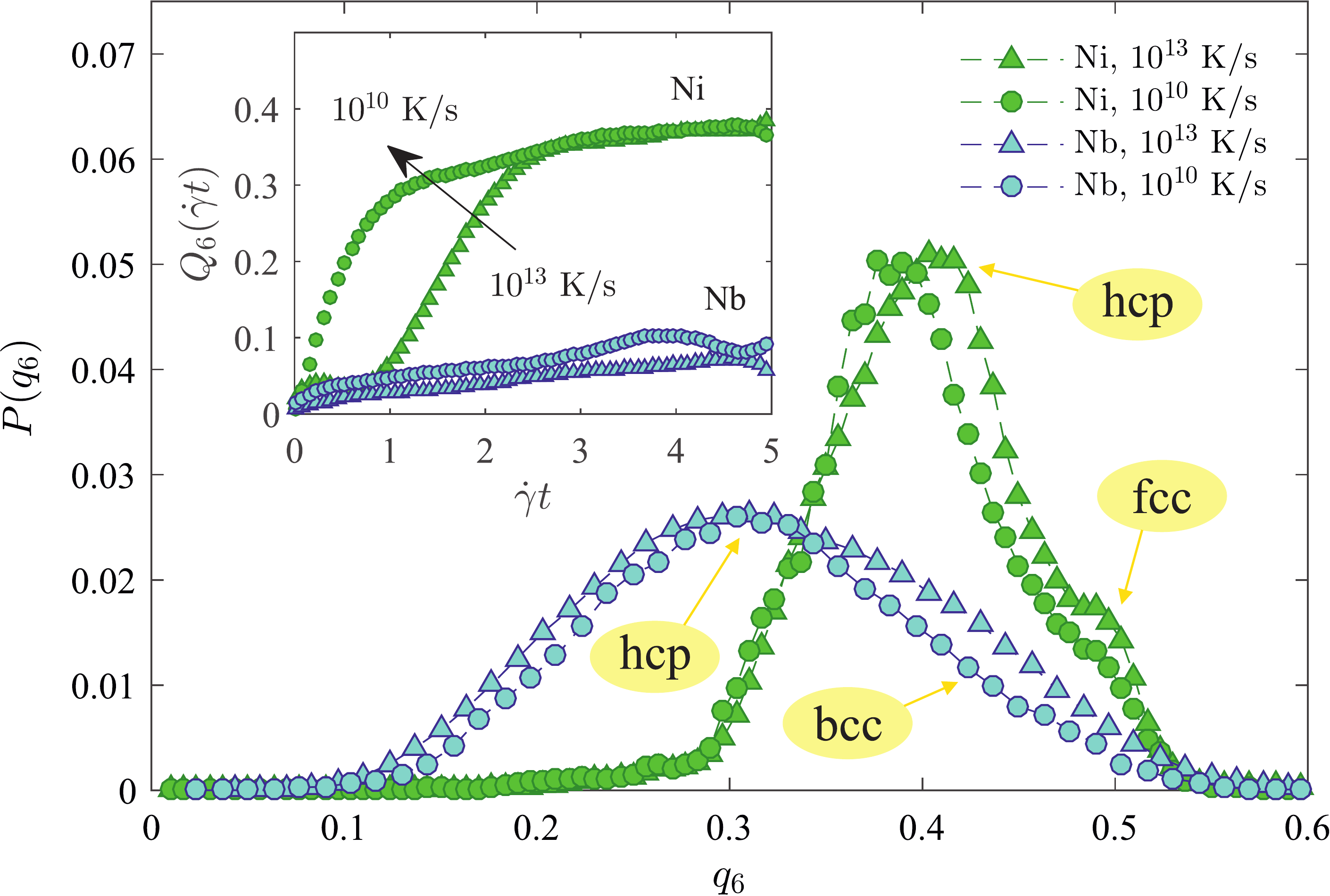}
	\caption{Distributions $P(q_{6})$ of crystalline Ni and Nb atoms obtained for the samples of considered BMG, which are prepared at different cooling rates. The curves are obtained at the shear strain $\dot{\gamma}t=5$ and at the pressure $p=600$~GPa. Inset: global orientational order parameter $Q_6$ as a function of the strain $\dot{\gamma}t$ calculated for the samples, which were cooled with different rates (for the case of $p=600$~GPa).}\label{fig_4}
\end{figure*}

It should be noted that the final crystal structure weakly depends on the cooling rate with which the sample has been quenched. For example, Figure~\ref{fig_4} shows the distribution of the order parameter $q_{6}$ calculated at the strain $\dot\gamma=5 $ and at the pressure $p=600$~GPa. The results are given for glassy samples prepared at the cooling rates $10^{10}$~K/s and $10^{13}$~K/s. Note that the realization of the cooling procedure with rates less than $10^{10}$~K/s in molecular dynamics simulations can take several months even when using supercomputers. Despite the significant difference in cooling rates, the shape of the $P(q_{q})$-distribution changes insignificantly and retains the positions of the main maxima and peaks corresponding to the fcc, hcp and bcc crystalline phases. On the other hand, the inset in Figure~\ref{fig_4} shows that the dependence of the global orientational order parameter $Q_{6}$ on the strain $\dot\gamma t$ calculated at the pressure $p=600$~GPa saturates faster at the cooling rate $10^{10}$~K/s. This result shows that the crystallization kinetics of the considered system depends on the cooling history: a slow cooling rate leads to a faster relaxation of the system into the crystalline phase. This result is in agreement with previous experimental and simulation studies of other BMG's~\cite{Vollmayr_Kob_1996,Borek_2014,Zhang_Ho_2015,Ryltsev_Klumov_2016}, where it was shown the positive effect of decreasing cooling rate on the crystallization kinetics as well as a weak dependence of the structural characteristics on the cooling procedure.  

\begin{figure*}[ht]
	\centering
	\includegraphics[width=15.0cm]{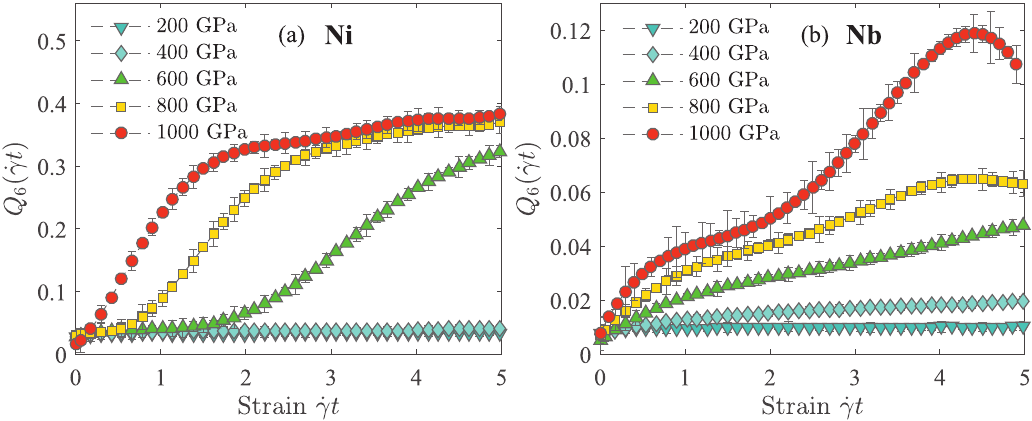}
	\caption{Global orientational order parameter $Q_{6}$ as a function of the strain $\dot{\gamma}t$: (a) for the crystal structure of Ni; (b) for the crystal structure of Nb.}\label{fig_5}
\end{figure*}

Figure~\ref{fig_5} shows the relationship between the global order parameter $Q_6$ and the strain $\dot{\gamma} t$ calculated at different pressures and for samples prepared at the cooling rate $10^{13}$~K/s. Note that the $Q_6$ is the most sensitive to the processes of structural ordering in comparison with other order parameters~\cite{Mickel_Kapfer_2013}. As seen from Figure~\ref{fig_5}, the growth kinetics of the Ni and Nb crystalline phases are very different. Namely, Ni atoms tend to order much rapidly than Nb atoms, and this tendency becomes more pronounced with increasing pressure. In the case of Ni, this is indicated by the rapid saturation of the $Q_6(\dot{\gamma} t)$-curves at pressures $p\geq600$~GPa, at which the parameter $Q_6$ takes the value $\approx0.4$ corresponding to the fcc/hcp phase. In the case of Nb, the value of the parameter $Q_6$ demonstrates a slower growth in comparison with the case of Ni atoms at the same pressures. Here, the $Q_6(\dot{\gamma} t)$-curves do not form a pronounced stable plateau, which indicates a slow ordering of Nb atoms.

\begin{figure*}[ht]
	\centering
	\includegraphics[width=14.0cm]{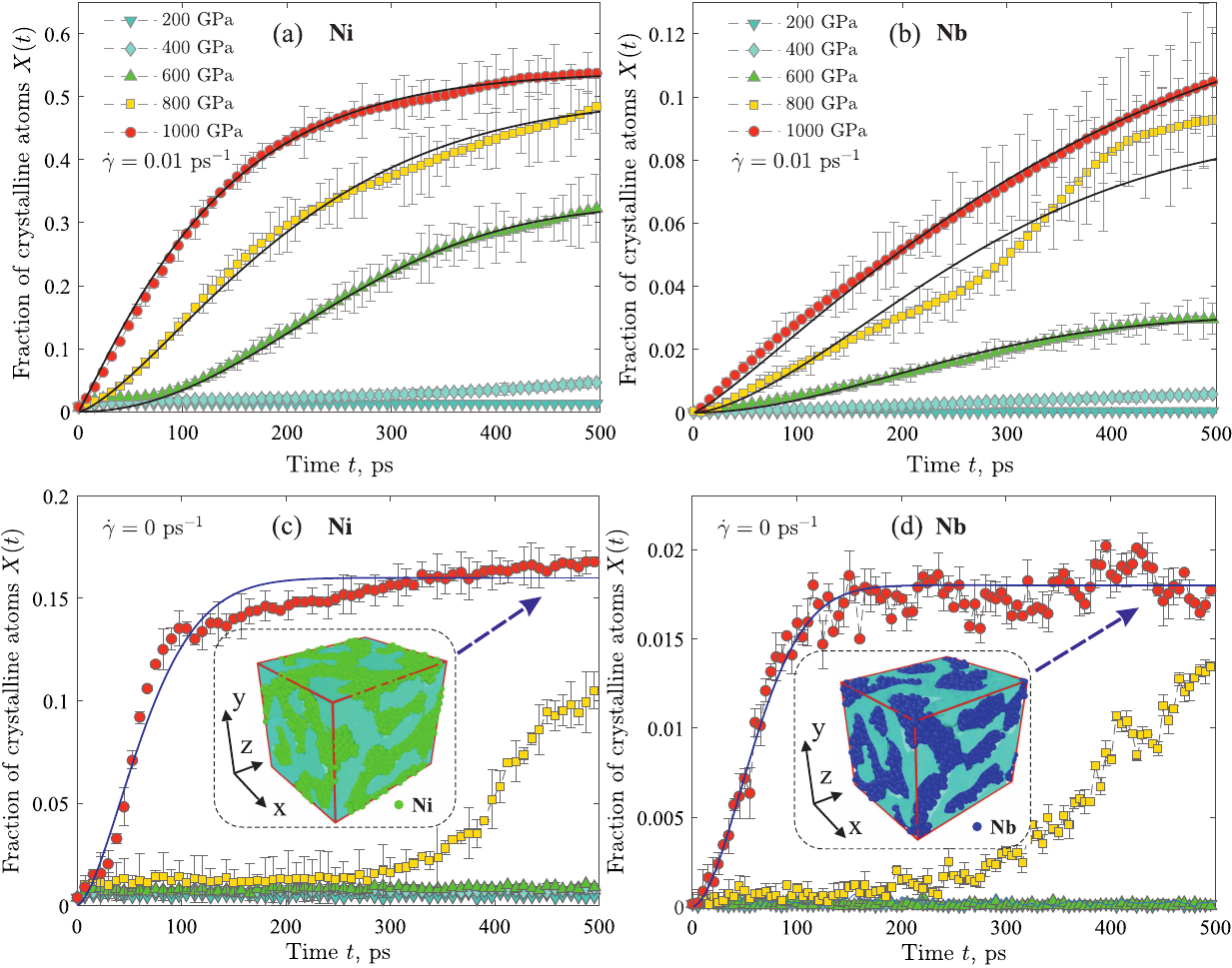}
	\caption{Fraction of the crystalline phase atoms $X$ as a function of the time $t$ computed for Ni and Nb atoms at different pressures (for samples cooled with the rate $10^{13}$~K/s): (a)--(b) at the shear rate $\dot\gamma = 0.01$~ps$^{-1}$ and (c)--(d) at zero shear rate $\dot\gamma = 0$~ps$^{-1}$. Solid lines are the results of Eq.~(\ref{eq_KJMAE}). Insets in panels (c) and (d) show the result of separation of Ni and Nb atoms in the system under the pressure $p=1000$~GPa.}\label{fig_6}
\end{figure*}

Figures~\ref{fig_6}(a) and~\ref{fig_6}(b) show the partial fraction of crystalline atoms $X\equiv n/N$ (where $n$ is the number of crystalline atoms, $N$ is the number of all atoms in the system) as a function of the time $t$ at different pressures. The fraction of the crystalline phase increases both with the increasing pressure $p$ and with the increasing strain $\dot\gamma t$. The results presented in Figures~\ref{fig_6}(a) and~\ref{fig_6}(b) also confirm that the crystalline Ni phase grows rapidly than the fraction of crystalline Nb atoms. This is mainly due to the low concentration of Nb atoms in the considered system. As a consequence, the formation of stable Nb crystallites is slowed down due to the relatively large distances between Nb atoms. External high pressure increases the local partial density of each component, which in combination with the shear initiates the crystallization.

For quantitative evaluation of phase transformation kinetics, we fit the calculated $X(t)$-curves with the exponential function
\begin{equation}\label{eq_KJMAE}
X_{\alpha}(t)=X_{m}^{(\alpha)}\left[1-e^{-K_{\alpha}t^{\xi}}\right],\,\,\alpha=\{\text{Ni},\,\text{Nb}\},
\end{equation}
which converts into the well-known Kolmogorov-Johnson-Mehl-Avrami-Evans (KJMAE) expression at $X_{m}^{(\alpha)}=1$~\cite{Kashchiev_2000,Mokshin_Barrat_2010,Lelito_2020,Yarullin_2020}. The parameter $X_{m}^{(\alpha)}$ defines the position of the plateau in the $t$-dependencies of the quantity $X_{\alpha}$. Note that this quantity is $X_{m}^{(\alpha)}=1$ in the case of crystallization of single-component systems and in the case of complete crystallization of a system~\cite{Lelito_2020}. The kinetic rate factor $K_{\alpha}$ [in units of $1/$ps$^{\xi}$] depends on the nucleation and crystal growth rates; $\xi$ is the so-called Avrami exponent. According to the KJMAE-theory, this exponent is $\xi\simeq3$ for the case of three-dimensional crystal growth. If the phase transformation occurs with a constant nucleation rate and the crystal growth is controlled by diffusion, then we have $\xi\simeq 2.5$. When the crystal growth is controlled only by the diffusion, the exponent $\xi$ takes the value close to $1.5$. Expression (\ref{eq_KJMAE}) correctly reproduces the $X(t)$-curves calculated at the pressures $p=600$, $800$ and $1000$~GPa [see Figures~\ref{fig_6}(a) and~\ref{fig_6}(b)]. The found values of the parameters of Eq.~(\ref{eq_KJMAE}) are given in Table~I.
\begin{table}[h]
	\begin{center}
		\small
		\caption{The values of the parameters of Eq.~(\ref{eq_KJMAE}) obtained by fitting the computed $X_{\alpha}(t)$-curves, where $\alpha=\{$Ni,~Nb$\}$. The values of the parameters $X_{m}^{(\alpha,\,\dot\gamma=0)}$, $K_{\alpha}^{(\dot\gamma=0)}$ and $\xi^{(\dot\gamma=0)}$ computed for the case of crystallization without shear, $\dot\gamma=0$.} 
		\begin{tabular}[t]{cccccccc}
			\hline
			$\alpha$ & p, GPa & $X_{m}^{(\alpha)}$ & $K_{\alpha}$, $\times10^{-4}$ps$^{-\xi}$ & $\xi$ & $X_{m}^{(\alpha,\,\dot\gamma=0)}$ & $K_{\alpha}^{(\dot\gamma=0)}$, $\times10^{-4}$ps$^{-\xi}$ & $\xi^{(\dot\gamma=0)}$ \\
			\hline
			& $1000$ & $0.54\pm0.02$ & $46.1\pm3.5$  & $1.1\pm0.1$ & $0.16\pm0.03$ & $9.8\pm0.7$ & $1.6\pm0.1$ \\
			Ni	& $800$  & $0.5\pm0.02$  & $5.12\pm0.4$   & $1.4\pm0.1$ & -- & -- & -- \\
			& $600$  & $0.33\pm0.02$ & $0.07\pm0.01$ & $2.1\pm0.2$ & -- & -- & -- \\
			\hline
			& $1000$ & $0.14\pm0.02$ & $7.96\pm0.6$  & $1.2\pm0.1$ & $0.018\pm0.002$ & $4.5\pm0.3$ & $1.8\pm0.1$ \\
			Nb	& $800$  & $0.09\pm0.01$ & $1.07\pm0.08$ & $1.6\pm0.1$ & -- & -- & -- \\
			& $600$  & $0.03\pm0.01$ & $0.02\pm0.01$ & $1.9\pm0.1$ & -- & -- & -- \\
			\hline
		\end{tabular}
	\end{center}\label{tab_1}
\end{table}

It was found that the exponent $\xi$ weakly depends on the crystalline phase composition and takes values from the range $\xi\in[1.1;\,2.1]$. Such small values confirm that the crystal growth is not uniform. The value of the exponent $\xi$ increases from ($1.1\pm0.1$) to ($2.1\pm0.1$) with decreasing pressure that indicates a possible transition to the classical nucleation scenario at low pressures. Moreover, the value of the kinetic rate factor $K_{\alpha}$ increases by several orders of magnitude with increasing pressure [see Table~I] that confirms the conclusion about ``explosive'' crystallization of the considered BMG at high pressures. For example, the growth of Ni crystals at the pressure $p=1000$~GPa occurs with the kinetic rate factor $K_{\text{Ni}}\simeq46\times10^{-4}$~ps$^{-1.1}$, while the crystalline Nb phase grows with the rate factor $K_{\text{Nb}}\simeq7.96\times10^{-4}$~ps$^{-1.2}$ at the same pressure. This difference in the values of the kinetic rate factor indicates that the crystallization of Ni$_{62}$Nb$_{38}$ occurs in two stages. At the first stage, the system with the crystalline Ni phase is formed intensively with a small amount of Nb crystals. The second stage is accompanied by an increasing proportion of Nb crystals, whereas the growth of crystalline Ni phase is slowing down. These conclusions are also confirmed by the results shown in Figures~\ref{fig_2} and~\ref{fig_6}.

To estimate the contribution of a pressure on the crystallization of the considered BMG, we have performed calculations without applying a shear, $\dot\gamma=0$. Figures~\ref{fig_6}(c) and~\ref{fig_6}(d) show a stable increase in the fraction of the crystalline phases Ni and Nb only at pressures above $600$~GPa. For example, at the pressure $p=1000$~GPa, the fraction of crystalline Ni and Nb atoms rapidly reaches a saturation that is accompanied by the formation of a pronounced plateau in the $X(t)$-curves. It is noteworthy that the formation of this plateau is preceded by the separation of Ni and Nb atoms to branched structures consisting of a mixture of crystalline and disordered parent phases [see insets in Figures~\ref{fig_6}(c) and~\ref{fig_6}(d)]. It follows from the position of the plateau that only $15$\% of Ni atoms and $1.8$\% of Nb atoms form the crystalline phase. For comparison, under shear strain and at the pressure $p=1000$~GPa, the plateau is reached at the fractions of crystalline Ni and Nb atoms $50$\% and $10$\%, respectively. This indicates the failure of the system to complete crystallization at the considered pressures and time scale. Approximation of the $X(t)$-curve by Eq.~(\ref{eq_KJMAE}) gives the kinetic velocity factor $K_{\alpha}^{(\dot\gamma=0)}=9.8\times10^{-4}$~ps$^{-\xi}$ (where $\xi=1.6$) for Ni atoms that is $\approx4.7$ times less than the case of crystallization under shear. In the case of Nb atoms, the kinetic velocity factor decreases only $\approx1.7$ times. In the absence of shear, the value of the parameter $\xi$ is greater than in the case of shear strain [see Table~\ref{tab_1}]. This is due to the fact that the growth of nuclei at zero shear rate occurs more uniformly in all directions. It should be noted that Eq.~(\ref{eq_KJMAE}) misdescribes the $X(t)$-curves at other pressures due to the poor crystallizability of the system. Thus, the ``explosive'' and complete crystallization of BMG in the considered time scale can be provided by applying shear deformation.

\section{Conclusions}

The results of the present work show that ultrahigh pressure and shear deformation promote the crystallization of Ni$_{62}$Nb$_{38}$, whereas this BMG forms a relatively stable amorphous structure under normal conditions. We reveal the presence of a threshold pressure, the excess of which triggers the structural ordering of the glassy system. Shear deformation does not lead to crystallization at pressures below the threshold. In this case, the crystallization of Ni$_{62}$Nb$_{38}$ proceeds through the separation of the system into two crystalline fractions. The kinetic rate factor of Ni crystallites rapidly increases with pressure, and this rate factor is several times higher than the growth rate factor of Nb crystals. On the one hand, these results show that the pressure $p$ is a key factor controlling BMG crystallization. On the other hand, it follows from the obtained results that the useful functional properties of BMG (for example, corrosion resistance, strength, hardness, and weak dependence of electrical resistance on temperature) directly related to the presence of a homogeneous amorphous structure can be lost under extreme conditions. The applied pressure and shear strain do not lead to the formation of BMG with higher-energy metastable state, which is also called rejuvenation effect~\cite{Ding_Zaccone_2019}. The reason for this can be the peculiarities of the interatomic interactions, which do not allow the formation of a high local energy barrier around the atom at high pressures. Note that the rejuvenation can lead to strain-hardening of some metallic glasses. This was demonstrated earlier by the example of ternary and quaternary systems based on zirconium and palladium~\cite{Pan_Greer_2020,Phan_Zaccone_2021}.

In this regard, it seems interesting to develop the results of the present study to solve the following problems:

(i) Crystalline separation and differences in the kinetic rate factors of the crystalline phases of Ni and Nb should impact the values of the main mechanical characteristics of Ni$_{62}$Nb$_{38}$. It would be interesting to test or predict the mechanical properties of this BMG under extreme conditions.

(ii) To study the phase diagram of Ni$_{62}$Nb$_{38}$ at high pressures.

(iii) Verification of the applicability of the crystallization conditions and the observed scenario of the phase transition (crystal separation, difference in crystal growth kinetics) in the case of other binary and ternary BMG's.

\section*{Acknowledgement}
\noindent This study is supported by the Russian Science Foundation (project No. 19-12-00022).

\section*{Appendix: Calculation of orientational order parameters}

According to Refs.~\cite{Steinhardt_1983,Wolde_Frenkel_1995}, the local orientational order parameters $q_{l}$ ($l=4,\,6,\,8$) are defined as follows
\begin{equation}\label{eq_A1}
q_{l}(\alpha,i)=\left[\frac{4\pi}{2l+1}\sum_{m=-l}^{l}\left|\bar{q}_{lm}(\alpha,i)\right|^{2}\right]^{1/2},\,\,i=1,2,...,N_{\alpha},
\end{equation}
where local structure around $i$th atom characterizes by
\begin{equation}\label{eq_A2}
\bar{q}_{lm}(\alpha,i)=\frac{1}{n_{\alpha,i}}\sum_{j=1}^{n_{\alpha,i}}Y_{lm}(\theta_{\alpha_i\alpha_j},\phi_{\alpha_i\alpha_j}).
\end{equation}
Here, $Y_{lm}(\theta_{\alpha_i\alpha_j},\phi_{\alpha_i\alpha_j})$ are the spherical harmonics ($\alpha=\{$Ni$,\,$Nb$\}$); the quantities $\theta_{\alpha_i\alpha_j}$ and $\phi_{\alpha_i\alpha_j}$ determine the polar and azimuthal angles, respectively; $n_{\alpha,i}$ is the number of nearest neighbors of $i$th atom. An atom $i$ is considered entering the crystalline phase if it forms an ordered structure with four or more $j$th atoms from the nearest environment. In this case, the atoms $i$ and $j$ belong to a common crystal cluster with the following correlation condition~\cite{Wolde_Frenkel_1995}:
\begin{equation}\label{eq_A3}
\left|\sum_{m=-6}^{6}\bar{q}_{6m}(\alpha,i)\bar{q}_{6m}^{*}(\alpha,j)\right|>0.5,
\end{equation}
where
\begin{equation}\label{eq_A4}
\bar{q}_{6m}^{*}(\alpha,j)=\frac{\bar{q}_{6m}(\alpha,j)}{\sqrt{\sum_{m=-6}^{6}\left|\bar{q}_{6m}(\alpha,j)\right|^{2}}}.
\end{equation} 
By calculating the local invariants $\bar{q}_{lm}(\alpha,i)$, we define the global orientational order parameters $Q_{l}(\alpha)$
\begin{equation}\label{eq_A5}
Q_{l}(\alpha)=\left(\frac{4\pi}{2l+1}\sum_{m=-l}^{l}\left|\bar{Q}_{lm}(\alpha)\right|^{2}\right)^{1/2},
\end{equation}
where
\begin{equation}\label{eq_A6}
\bar{Q}_{lm}(\alpha)=\frac{\sum_{i=1}^{N_{\alpha}}n_{\alpha,i}\cdot\bar{q}_{lm}(\alpha,i)}{\sum_{i=1}^{N_{\alpha}}n_{\alpha,i}}.
\end{equation}
In the case of liquids and amorphous solids, the order parameters $Q_{l}$ ($l=4,\,6,\,8$) take values close to zero. For the crystalline phase, these parameters take specific values that characterize the type of crystal lattices~\cite{Mickel_Kapfer_2013}.

\bibliographystyle{unsrt}

\begin{thebibliography}{57}


\bibitem{Dubrovinskaia_2016}
N. Dubrovinskaia, L. Dubrovinsky, N.A. Solopova, A. Abakumov, S. Turner, M. Hanfland, E. Bykova, M. Bykov, C. Prescher,  V.B. Prakapenka, S. Petitgirard, I. Chuvashova, B. Gasharova, Y.-L. Mathis, P. Ershov, I. Snigireva, A. Snigirev, Terapascal static pressure generation with ultrahigh yield strength nanodiamond, Sci. Adv. 2 (2016) e1600341. https://doi.org/10.1126/sciadv.1600341

\bibitem{Zhang_Ma_2017}
L. Zhang, Y. Wang, J. Lv, Y. Ma, Materials discovery at high pressures, Nat. Rev. Mater. 2 (2017) 17005. https://doi.org/10.1038/natrevmats.2017.5

\bibitem{Burrage_Perreault_2019}
K.C. Burrage, C.S. Perreault, E.K. Moss, J.S. Pigott, B.T. Sturtevant, J.S. Smith, N. Velisavljevic, Y.K. Vohra, Ultrahigh pressure equation of state of tantalum to 310~GPa, High Press. Res. 39 (2019) 489--498. https://doi.org/10.1080/08957959.2019.1641203

\bibitem{Glezer_2019}
A.M. Glezer, D.V. Louzguine-Luzgin, I.A. Khriplivets, R.V. Sundeev, D.V. Gunderov, A.I. Bazlov, Yu.S. Pogozhev, Effect of high-pressure torsion on the tendency to plastic flow in bulk amorphous alloys based on Zr, Mater. Lett. 256 (2019) 126631. https://doi.org/10.1016/j.matlet.2019.126631

\bibitem{Dubrovinsky_2015}
L. Dubrovinsky, N. Dubrovinskaia, E. Bykova, M. Bykov, V. Prakapenka, C. Prescher, K. Glazyrin, H.-P. Liermann, M. Hanfland, M. Ekholm, Q. Feng, L.V. Pourovskii, M.I. Katsnelson, J. M.Wills, and I.A. Abrikosov, The most incompressible metal osmium at static pressures above 750 gigapascals, Nature 525 (2015) 226--229. https://doi.org/10.1038/nature14681

\bibitem{Dewaele_Loubeyre_2018}
A. Dewaele, P. Loubeyre, F. Occelli, O. Marie, M. Mezouar, Toroidal diamond anvil cell for detailed measurements under extreme static pressures, Nat. Commun. 9 (2018) 2913. https://doi.org/10.1038/s41467-018-05294-2

\bibitem{Brazhkin_2020}
V.V. Brazhkin, Ultrahard nanomaterials: myths and reality, Phys.-Usp. 63 (2020) 523--544. https://doi.org/10.3367/UFNe.2019.07.038635

\bibitem{Brazhkin_Tsiok_2017}
V.V. Brazhkin, E. Bychkov, O.B. Tsiok, High-precision measurements of the compressibility and the electrical resistivity of bulk g-As$_2$Te$_3$ glasses at a hydrostatic pressure up to $8.5$~GPa, J. Exp. Theor. Phys. 125 (2017) 451--464. https://doi.org/10.1134/S1063776117080155 

\bibitem{Xing_Jiang_2002}
P.F. Xing, Y. Zhuang, W.H. Wang, L. Gerward, and J. Jiang, Glass transition, crystallization kinetics and pressure effect on crystallization of ZrNbCuNiBe bulk metallic glass, J. Appl. Phys. 91 (2002) 4956--4960. https://doi.org/10.1063/1.1461892

\bibitem{Sosso_2016}
G.C. Sosso, J. Chen, S.J. Cox, M. Fitzner, P. Pedevilla, A. Zen, and A. Michaelides, Crystal nucleation in liquids: open questions and future challenges in molecular dynamics simulations, Chem. Rev. 116 (2016) 7078--7116. https://doi.org/10.1021/acs.chemrev.5b00744

\bibitem{Mo_Liu_2017}
J. Mo, H. Liu, Y. Zhang, M. Wang, L. Zhang, B. Liu, W. Yang, Effects of pressure on structure and mechanical property in monatomic metallic glass, J. Non-Cryst. Solids 464 (2017) 1--4. https://doi.org/10.1016/j.jnoncrysol.2017.03.013

\bibitem{Watanauki_Shobu_2006}
T. Watanauki, A. Machida, T. Lkeda, K. Aoki, H. Kaneko,
T. Shobu, T.J. Sato and A.P. Tsai, Pressure-induced phase transitions in the Cd-Yb periodic approximant to a quasicrystal, Phys. Rev. Lett. 96 (2006) 105702. https://doi.org/10.1103/PhysRevLett.96.105702

\bibitem{Galimzyanov_Mokshin_2018}
B.N. Galimzyanov, D.T. Yarullin, A.V. Mokshin, Change in the crystallization features of supercooled liquid metal with an increase in the supercooling level, JETP Letters 107 (2018) 629--634. https://doi.org/10.1134/S0021364018100089

\bibitem{Galimzyanov_AM_2019}
B.N. Galimzyanov, D.T. Yarullin, A.V. Mokshin, Structure and morphology of crystalline nuclei arising in a crystallizing liquid metallic film, Acta Mater. 169 (2019) 184--192. https://doi.org/10.1016/j.actamat.2019.03.009

\bibitem{Soignard_Amin_2008}
E. Soignard, S.A. Amin, Q. Mei, C.J. Benmore, and J.L. Yarger, High-pressure behavior of As$_2$O$_3$: Amorphous-amorphous and crystalline-amorphous transitions, Phys. Rev. B 77 (2008) 144113. https://doi.org/10.1103/PhysRevB.77.144113

\bibitem{Wang_Yao_2001}
W.H. Wang, R.J. Wang, D.Y. Dai, D.Q. Zhao, M.X. Pan, and Y.S. Yao, Pressure-induced amorphization of ZrTiCuNiBe bulk glass-forming alloy, Appl. Phys. Lett. 79 (2001) 1106--1108. https://doi.org/10.1063/1.1396321

\bibitem{Soignard_Brazhkin2020}
E. Soignard, O.B. Tsiok, A.S. Tverjanovich, A. Bytchkov, A. Sokolov, V.V. Brazhkin, C.J. Benmore, and E. Bychkov, Pressure-driven chemical disorder in glassy As$_2$S$_3$ up to $14.7$~GPa, postdensification effects, and applications in materials design, J. Phys. Chem. B 124 (2020) 430--442. https://doi.org/10.1021/acs.jpcb.9b10465

\bibitem{Halim_Ma_2021}
Q. Halim, N.A.N. Mohamed, M.R.M. Rejab, W.N.W.A. Naim and Q. Ma, Metallic glass properties, processing method and development perspective: a review, Int. J. Adv. Manuf. Technol. 112 (2021) 1231--1258. https://doi.org/10.1007/s00170-020-06515-z

\bibitem{Li_Adv_2020}
Z. Li, Z. Huang, F. Sun, X. Li, J. Ma, Forming of metallic glasses: mechanisms and processes, Mater. Today Adv. 7 (2020) 1000772. https://doi.org/10.1016/j.mtadv.2020.100077

\bibitem{Xia_Li_2006}
L. Xia, W.H. Li, S.S. Fang, B.C. Wei and Y.D. Dong, Binary Ni-Nb bulk metallic glasses, J. Appl. Phys. 99 (2006) 026103. https://doi.org/10.1063/1.2158130

\bibitem{Suryanarayana_2018}
C. Suryanarayana, Phase formation under non-equilibrium processing conditions: rapid solidification processing and mechanical alloying, J. Mater. Sci. 53 (2018) 13364--13379. https://doi.org/10.1007/s10853-018-2197-4

\bibitem{Massalski_Okamoto_1990}
T.B. Massalski, H. Okamoto, P.R. Subramanian, L. Kacprzak, Binary alloy phase diagrams, 2nd edn. ASM International, Materials Park, 1990.

\bibitem{Lesz_Dercz_2016}
S. Lesz, G. Dercz, Study on crystallization phenomenon and thermal stability of binary Ni-Nb amorphous alloy, J. Therm. Anal. Calorim. 126 (2016) 19--26. https://doi.org/10.1007/s10973-016-5786-y

\bibitem{Liu_Chen_2008}
X.J. Liu, G.L. Chen, H.Y. Hou, X. Hui, K.F. Yao, Z.P. Lu, C.T. Liu, Atomistic mechanism for nanocrystallization of metallic glasses, Acta Mater. 56 (2008) 2760--2769. https://doi.org/10.1016/j.actamat.2008.02.019

\bibitem{Carter_Williams_2009}
C.B. Carter, D.B. Williams, Transmission Electron Microscopy: A Textbook for Materials Science, Springer, New York, 2009.

\bibitem{Ediger_Harrowell_2012}
M.D. Ediger, P. Harrowell, Perspective: supercooled liquids and glasses, J. Chem. Phys. 137 (2012) 080901. https://doi.org/10.1063/1.4747326

\bibitem{Gleiter_2000}
H. Gleiter, Nanostructured materials: basic concepts and microstructure, Acta Mater. 48 (2000) 1--29. https://doi.org/10.1016/S1359-6454(99)00285-2

\bibitem{Myerson_2002}
A. Myerson, Handbook of Industrial Crystallization, second ed., Butterworth Heinemann, London, 2002.

\bibitem{Kelton_2016}
Y. Zhang, R. Ashcraft, M.I. Mendelev, C.Z. Wang, and K.F. Kelton, Experimental and molecular dynamics simulation study of structure of liquid and amorphous Ni$_{62}$Nb$_{38}$ alloy, J. Chem. Phys. 145 (2016) 204505. https://doi.org/10.1063/1.4968212

\bibitem{Canales_Padro_1992}
M. Canales, J. A. Padro', On the Bond-angle distributions in liquids and liquid solutions, Mol. Simul. 8 (1992) 335--344. https://doi.org/10.1080/08927029208022488

\bibitem{Finnis_Sinclair_1984}
M.W. Finnis and J.E. Sinclair, A simple empirical N-body potential for transition metals, Philos. Mag. A 50, (1984) 45--55. https://doi.org/10.1080/01418618408244210

\bibitem{Xu_Jiang_2017}
T.D. Xu, X.D. Wang, H. Zhang, Q.P. Cao, D.X. Zhang, and J.Z. Jiang, Structural evolution and atomic dynamics in Ni-Nb metallic glasses: A molecular dynamics study, J. Chem. Phys. 147 (2017) 144503. https://doi.org/10.1063/1.4995006

\bibitem{Mokshin_Barrat_2013}
A.V. Mokshin, B.N. Galimzyanov, and J.-L. Barrat, Extension of classical nucleation theory for uniformly sheared systems, Phys. Rev. E 87 (2013) 062307. https://doi.org/10.1103/PhysRevE.87.062307

\bibitem{Chang_Zhou_2017}
L. Chang, C.-Y. Zhou, L.-L. Wen, J. Li, X.-H. He, Molecular dynamics study of strain rate effects on tensile behavior ofsingle crystal titanium nanowire, Comput. Mater. Sci. 128 (2017) 348--358. https://doi.org/10.1016/j.commatsci.2016.11.034

\bibitem{Sha_2018}
Z.-D. Sha, Q.-X. Pei, K. Zhou, Z. Dong, Y.-W. Zhang, Temperature and strain-rate dependent mechanical properties of single-layer borophene, Extreme Mech. Lett. 19 (2018) 39--45. https://doi.org/10.1016/j.eml.2017.12.008

\bibitem{Steinhardt_1983}
P. Steinhardt, D. Nelson, and M. Ronchetti, Bond-orientational order in liquids and glasses, Phys. Rev. B. 28 (1983) 784--805. https://doi.org/10.1103/PhysRevB.28.784

\bibitem{Wolde_Frenkel_1995}
P. ten Wolde, M. Ruiz-Montero, and D. Frenkel, Numerical evidence for bcc ordering at the surface of a critical fcc nucleus, Phys. Rev. Lett. 75 (1995) 2714--2717. https://doi.org/10.1103/PhysRevLett.75.2714

\bibitem{Mickel_Kapfer_2013}
W. Mickel, S.C. Kapfer, G.E. Schruder-Turk, K. Mecke, Shortcomings of the bond orientational order parameters for the analysis of disordered particulate matter, J. Chem. Phys. 138 (2013) 044501. https://doi.org/10.1063/1.4774084

\bibitem{Mokshin_Galimzyanov_2017}
A.V. Mokshin, B.N. Galimzyanov, Kinetics of the crystalline nuclei growth in glassy systems, Phys. Chem. Chem. Phys. 19 (2017) 11340--11353. https://doi.org/10.1039/C7CP00879A

\bibitem{Zhang_Duan_2019}
D. Zhang, G. Lu, D. Ruan, Q. Fei, W. Duan, Quasi-static combined compression-shear crushing of honeycombs: An experimental study, Materials \& Design 167 (2019) 107632. https://doi.org/10.1016/j.matdes.2019.107632

\bibitem{Mura_Zaccone_2016}
F. Mura and A. Zaccone, Effects of shear flow on phase nucleation and crystallization, Phys. Rev. E 93 (2016) 042803. https://doi.org/10.1103/PhysRevE.93.042803

\bibitem{Galimzyanov_JR_2018}
B.N. Galimzyanov, A.V. Mokshin, Morphology of critically sized crystalline nuclei at shear-induced crystal nucleation in amorphous solid, J. Rheol. 62 (2018) 265--275. https://doi.org/10.1122/1.5003238

\bibitem{Kashchiev_2000}
D. Kashchiev, Nucleation: Basic Theory with Applications, Butterworth Heinemann, Oxford, UK, 2000.

\bibitem{Kelton_Greer_2010}
K.F. Kelton, A.L. Greer, Nucleation in Condensed Matter, Elsevier, Amsterdam, 2010.

\bibitem{Kalikmanov_2012}
V.I. Kalikmanov, Nucleation Theory, Lecture Notes in Physics, Springer, New York, 2012.

\bibitem{Schmelzer_Abyzov_2018}
J.W.P. Schmelzer, A.S. Abyzov, Crystallization of glass-forming melts: new answers to old questions, J. Non-Cryst. Solids 501 (2018) 11--20. https://doi.org/10.1016/j.jnoncrysol.2017.11.047

\bibitem{Karthika_Kalaichelvi_2016}
S. Karthika, T.K. Radhakrishnan, and P. Kalaichelvi, A review of classical and nonclassical nucleation theories, Cryst. Growth Des. 16 (2016) 6663--6681. https://doi.org/10.1021/acs.cgd.6b00794

\bibitem{Skripov_1979}
V.P. Skripov, A.V. Skripov, Spinodal decomposition (phase transitions via unstable states), Sov. Phys. Usp. 22 (1979) 389--410. https://doi.org/10.1070/PU1979v022n06ABEH005571

\bibitem{Igolnikov_Skripov_2021}
A.A. Igolnikov, S.B. Rutin, P.V. Skripov, Short-term measurements in thermally-induced unstable states of mixtures with LCST, Thermochim. Acta 695 (2021) 178815. https://doi.org/10.1016/j.tca.2020.178815

\bibitem{Yang_Tang_2020}
K. Yang, M. Tang, Three-dimensional phase evolution and stress-induced non-uniform Li intercalation behavior in lithium iron phosphate, J. Mater. Chem. A 8 (2020) 3060--3070. https://doi.org/10.1039/C9TA11697D

\bibitem{Rudraraju_2016}
S. Rudraraju, A. Van der Ven, K. Garikipati, Mechanochemical spinodal decomposition: a phenomenological theory of phase transformations in multi-component, crystalline solids, npj Comput. Mater. 2 (2016) 16012. https://doi.org/10.1038/npjcompumats.2016.12

\bibitem{Bachmaier_2015}
A. Bachmaier, M. Pfaff, M. Stolpe, H. Aboulfadl, C. Motz, Phase separation of a supersaturated nanocrystalline Cu-Co alloy and its influence on thermal stability, Acta Mater. 96 (2015) 269--283. https://doi.org/10.1016/j.actamat.2015.05.053

\bibitem{Sharma_Hickman_2018}
A. Sharma, J. Hickman, N. Gazit, E. Rabkin, Y. Mishin, Nickel nanoparticles set a new record of strength, Nat. Commun. 9 (2018) 4102. https://doi.org/10.1038/s41467-018-06575-6

\bibitem{Bollinger_White_2011}
R.K. Bollinger, B.D. White, J.J. Neumeier, H.R.Z. Sandim, Y. Suzuki, C.A.M. dos Santos, R. Avci, A. Migliori, and J.B. Betts, Observation of a martensitic structural distortion in V, Nb, and Ta, Phys. Rev. Lett. 107 (2011) 075503. https://doi.org/10.1103/PhysRevLett.107.075503

\bibitem{Vollmayr_Kob_1996}
K. Vollmayr, W. Kob, and K. Binder, How do the properties of a glass depend on the cooling rate? A computer simulation study of a Lennard‐Jones system, J. Chem. Phys. 105 (1996) 4714. https://doi.org/10.1063/1.472326

\bibitem{Borek_2014}
M. Krupi\'nski, B. Krupi\'nska, K. Labisz, Z. Rdzawski, W. Borek, Influence of cooling rate on crystallisation kinetics
on microstructure of cast zinc alloys, J. Therm. Anal. Calorim. 118 (2014) 1361. https://doi.org/10.1007/s10973-014-4174-8

\bibitem{Zhang_Ho_2015}
Y. Zhang, F. Zhang, C.Z. Wang, M.I. Mendelev, M.J. Kramer, and K.M. Ho, Cooling rates dependence of medium-range order development in Cu$_{64.5}$Zr$_{35.5}$ metallic glass, Phys. Rev. B 91 (2015) 064105. https://doi.org/10.1103/PhysRevB.91.064105

\bibitem{Ryltsev_Klumov_2016}
R.E. Ryltsev, B.A. Klumov, N.M. Chtchelkatchev, and K.Yu. Shunyaev, Cooling rate dependence of simulated Cu$_{64.5}$Zr$_{35.5}$ metallic glass structure, J. Chem. Phys. 145 (2016) 034506. https://doi.org/10.1063/1.4958631

\bibitem{Mokshin_Barrat_2010}
A.V. Mokshin and J.-L. Barrat, Crystal nucleation and cluster-growth kinetics in a model glass under shear, Phys. Rev. E 82 (2010) 021505. https://doi.org/10.1103/PhysRevE.82.021505

\bibitem{Lelito_2020}
J. Lelito, Crystallization Kinetics Analysis of the Amorphouse Mg$_{72}$Zn$_{24}$Ca$_{4}$ alloy at the isothermal annealing temperature of $507$~K, Materials 13 (2020) 2815. https://doi.org/10.3390/ma13122815

\bibitem{Yarullin_2020}
D.T. Yarullin, B.N. Galimzyanov, A.V. Mokshin, Direct evaluation of attachment and detachment rate factors of atoms in crystallizing supercooled liquids, J. Chem. Phys. 152 (2020) 224501. https://doi.org/10.1063/5.0007378

\bibitem{Ding_Zaccone_2019}
G. Ding, C. Li, A. Zaccone, W.H. Wang, H.C. Lei, F. Jiang, Z. Ling, M.Q. Jiang, Ultrafast extreme rejuvenation of metallic glasses by shock compression, Sci. Adv. 5 (2019) eaaw6249. 
https://doi.org/10.1126/sciadv.aaw6249

\bibitem{Pan_Greer_2020}
J. Pan, Yu. P. Ivanov, W.H. Zhou, Y. Li, A.L. Greer, Strain-hardening and suppression of shearbanding in rejuvenated bulk metallic glass, Nature 578 (2020) 559. https://doi.org/10.1038/s41586-020-2016-3

\bibitem{Phan_Zaccone_2021}
A.D. Phan, A. Zaccone, V.D. Lam, and K. Wakabayashi, Theory of Pressure-Induced Rejuvenation and Strain Hardening in Metallic Glasses, Phys. Rev. Lett. 126 (2021) 025502. https://doi.org/10.1103/PhysRevLett.126.025502


\end{thebibliography}

\end{document}